\documentclass[
 reprint,
 amsmath,amssymb,
 aps,
 prl]{revtex4-2}

\usepackage{graphicx}
\usepackage{dcolumn}
\usepackage{bm}
\usepackage[many]{tcolorbox}
\usepackage{lipsum}
\usepackage{amsmath,amssymb,cancel}

\usepackage{shuffle}
\usepackage{graphbox}
\usepackage{environ}
\usepackage{physics}
\usepackage{tabularx}
\usetikzlibrary{shapes.geometric}
\usepackage{asymptote}
\usepackage{slashed}
\usepackage{float}
\usepackage{subfig}
\usepackage{cleveref}

\newcommand*{\shifttext}[2]{%
  \settowidth{\@tempdima}{#2}%
  \makebox[\@tempdima]{\hspace*{#1}#2}%
}

\newcommand{\trmax}{\mathrm{tr_{max}}}
\newcommand{\trv}{\mathrm{tr}_{V}}
\newcommand{\trs}{\mathrm{tr}_{S}}

\newcommand{\fwl}{\mathrm{F.L.}}

\begin{document}

\title{On one-loop amplitudes in gauge theories}

\author{Qu Cao$^{2,5}$}
\email{qucao@zju.edu.cn}
\author{Jin Dong$^{2,4,6}$}
\email{dongjin@itp.ac.cn}
\author{Song He$^{1,2,3}$}
\email{songhe@itp.ac.cn}
\author{Fan Zhu$^{1,2,4}$}
\email{zhufan22@mails.ucas.ac.cn}
\affiliation{
$^{1}$School of Fundamental Physics and Mathematical Sciences, Hangzhou Institute for Advanced Study, UCAS \& ICTP-AP, Hangzhou, 310024, China\\
$^{2}$CAS Key Laboratory of Theoretical Physics, Institute of Theoretical Physics, Chinese Academy of Sciences, Beijing 100190, China \\
$^{3}$Peng Huanwu Center for Fundamental Theory, Hefei, Anhui 230026, P. R. China\\
$^{4}$School of Physical Sciences, University of Chinese Academy of Sciences, No.19A Yuquan Road, Beijing 100049, China\\
$^{5}$Zhejiang Institute of Modern Physics, Department of Physics, Zhejiang University, Hangzhou 310027, China \\
$^{6}$Department of Physics and Astronomy,
Uppsala University, 75120 Uppsala, Sweden
}

\begin{abstract}
We propose a new ``universal expansion" for one-loop amplitudes with arbitrary number of gluons in $D$ dimensions, which holds for general gauge theories with gluons/fermions/scalars in the loop, including pure and supersymmetric Yang-Mills theories. It expresses the $n$-gluon amplitudes as a linear combination of universal scalar-loop amplitudes with $n{-}m$ gluons and $m$ scalars, multiplied by gauge-invariant building blocks (defined for general gauge theories); the integrands of these scalar-loop amplitudes are given in terms of tree-level objects attached to the scalar loop, or by differential operators acting on the most important part which is proportional to $D$ (with $m=0$). We present closed-formula for these one-loop integrands and prove them by showing that the single cuts are correctly reproduced by the gluing of an additional pair of gluons (fermions/scalars) in the forward limit, plus $n$ gluons in a tree amplitude. 
\end{abstract}

\maketitle


\section{Introduction}
Recent years have witnessed enormous progress in the study of scattering amplitudes in quantum field theory and string theory~\cite{Elvang:2013cua,Henn:2014yza,Travaglini:2022uwo,Mafra:2022wml}. Not only have we seen new structures and new relations in gauge theories and gravity such as color-kinematics duality and the double copy construction~\cite{Bern:2008qj,Bern:2010ue,Bern:2019prr} (see {\it e.g.}~\cite{Carrasco:2023qgz, Edison:2023ulf, Li:2023akg,Li:2024zby,Bern:2024vqs} for recent developments), but there is even a new formalism for all-loop amplitudes in various colored theories including Yang-Mills theory~\cite{Arkani-Hamed:2023lbd,Arkani-Hamed:2023mvg,Arkani-Hamed:2023swr,Arkani-Hamed:2023jry,Arkani-Hamed:2024nhp}. In this letter, we report on new results in the study of loop amplitudes in general gauge theories in $D$ dimensions by combining two lines of research: the ``surface" approach to Yang-Mills loop integrands~\cite{Arkani-Hamed:2023jry, Arkani-Hamed:2024vna} and ideas related to double copy as well as worldsheet formulations such as Cachazo-He-Yuan (CHY) formulas~\cite{Cachazo:2013gna,Cachazo:2013iea,Cachazo:2013hca,Cachazo:2014xea}, which among other things produce closed-formulas for tree amplitudes with $n$ gluons and an additional pair of gluons/fermions/scalars~\cite{Dong:2021qai,Edison:2020uzf}. We will focus on one-loop amplitudes with arbitrary number of gluons in general gauge theories where fermions/scalars can also run in the loop. As reported in~\cite{Arkani-Hamed:2024tzl}, it is straightforward to reconstruct the one-loop integrands with $n$ gluons in pure Yang-Mills(YM) theory from the single cuts, which are in turn given by the forward limit of tree amplitudes with $n{+}2$ gluons. Remarkably, the notorious issue of divergences associated with such forward limits was canonically regulated by the ``surface kinematics" in~\cite{Arkani-Hamed:2024tzl}. On the other hand, it remains an important open question how to include matters such as fermions and scalars in this formalism, and as a first step we consider including them in the loop which amounts to taking forward limit of such an additional pair. Note that CHY formulas and the underlying ambitwistor string theories produce loop integrands with linear propagators~\cite{He:2015yua,Cachazo:2015aol,Geyer:2015jch,He:2016mzd,He:2017spx,Geyer:2017ela,Edison:2020uzf} whose single-cuts are naturally associated with ``forward limits" of trees considered in~\cite{Caron-Huot:2010fvq}. The strategy we adopt here is very different: with the help of surfaceology, by construction the Yang-Mills integrand reconstructed from single-cuts must have correct propagators which are associated with curves on the surface. The key new ingredient is a general formula for forward limit in general gauge theories~\cite{Edison:2020uzf,Dong:2021qai}, which played an important role for ``bootstraping" one-loop kinematic numerators, satisfying color-kinematics duality (thus they give corresponding gravity integrands via double copy)~\cite{Bern:2008qj,Bern:2010ue,Bern:2019prr}, {\it e.g.} up to $n=7$ in maximally supersymmetric Yang-Mills theory~\cite{Edison:2022jln}. By combining the two approaches, we obtain a new formula for one-loop $n$-gluon amplitudes in general gauge theories. 

We will derive the explicit formula of one-loop $n$-gluon integrands in Yang-Mills theory by showing that its single cuts agree with forward limits of trees with $n{+}2$ gluons (written in universal expansion of~\cite{Dong:2021qai}), which involves mixed amplitudes with gluons and at least two scalars (forming a scalar line); it is exactly the gluing of these two scalars that give the scalar loop, which allows us to write one-loop integrands in terms of scalar-loop amplitudes. 
From here the extension to general gauge theories becomes straightforward since such forward limits involving $n$ gluons and a pair of fermions/scalars are provided in~\cite{Edison:2020uzf} using the same mixed amplitudes. The upshot is that the one-loop expansion we obtain is also universal: we have exactly the same scalar-loop mixed amplitudes but different gauge-invariant prefactors which know about the species of particles running in the loop.

\section{A review of tree and one-loop amplitudes}\label{sec:review}
\paragraph{The universal expansion and forward limits}
Let us quickly review the universal expansion for tree amplitudes in gauge theories, which is a particularly nice form suited for taking the forward limit. As studied in~\cite{Dong:2021qai}, gluon tree amplitudes can be written as a combination of certain prefactors multiplied by simpler tree amplitudes with at least two external scalar. The key idea is that the forward limit acts trivially on both the prefactors and these tree amplitudes, transforming the former into distinct one-loop prefactors for different theories and the latter into universal scalar-loop amplitudes. Crucially, our reconstruction produces the integrand with propagators quadratic in the loop momentum, which is different from those in~\cite{He:2015yua,Cachazo:2015aol,Geyer:2015jch,He:2016mzd,He:2017spx,Geyer:2017ela,Edison:2020uzf}.

To begin with, let us introduce the universal expansion for $(n+2)-$point tree-level pure YM amplitudes:
\begin{equation} \label{eq:expansion}
\begin{aligned}
    &{\cal A}^\text{YM}(1,2,\ldots,i-1,-,+,i,\ldots,n)\\
    =& \sum_{m,\alpha\in S_m} \epsilon_- \cdot f_{\alpha_1} \cdot f_{\alpha_2} 
    \cdots f_{\alpha_m} \cdot \epsilon_+ \\
    & \times {\cal A}^\text{YMS}(-,\alpha,+|1,2,\ldots,i-1,-,+,i,\ldots,n)\,,
\end{aligned}
\end{equation}
where we have singled out two special legs $-,+$ and sum over all ordered subsets of $\{1,2,\ldots,n\}$ represented by $\alpha$ (with $|\alpha|=m$). 
The above expansion involves prefactor that manifests the gauge invariant of all particles in $\alpha$ via the linearized field strength:  $f_i^{\mu \nu}=k_i^\mu \epsilon_i^\nu -k_i^\nu \epsilon_i^\mu$, and each prefactor is multiplied by a Yang-Mills-scalar theory(YMS) amplitude with gluons $\Bar{\alpha}$ (the compliment of $\alpha$) and bi-adjoint $\phi^3$ scalars where the second color ordering is denoted by $(-,\alpha,+)$ (the first color ordering is the canonical ordering). This expansion was originally discovered using the CHY formulae~\cite{Lam:2016tlk,Fu:2017uzt,Du:2017kpo} and was later realized as a consequence of the fundamental principals in \cite{Dong:2021qai}. Remarkably, it is shown in~\cite{Edison:2020uzf} that based on the computation in the $-1/2,-3/2$ ghost picture for two fermions leg, the tree amplitudes in the supersymmetric Yang-Mills theory(SYM) exhibit a similar expansion:
\begin{equation} \label{eq:expansion2}
\begin{aligned}
    &{\cal A}^\text{SYM}(1,2,\ldots,i-1,-^f,+^f,i,\ldots,n)\\
    =& \sum_{m,\alpha\in S_m} \bar{\chi}_-  \slashed{f}_{\alpha_1}  \slashed{f}_{\alpha_2}
    \cdots \slashed{f}_{\alpha_m}  \xi_+ \\
    & \times {\cal A}^\text{YMS}(-,\alpha,+|1,2,\ldots,i-1,-,+,i,\ldots,n),
\end{aligned}
\end{equation}
where we have introduced the spinor wavefunction $\bar{\chi}_-$ and dual spinor wavefunction $\xi_+=\frac{\slashed{q} \chi_+}{2 q \cdot k_+}$ with $q$ to be a null reference momentum. The above expression involves Dirac gamma matrices that satisfy the Clifford algebra, $\Gamma^\mu \Gamma^\nu + \Gamma^\nu \Gamma^\mu = 2\eta^{\mu \nu}$. We define their contraction with the linearized field strength as $\slashed{f_i} = \frac{1}{4} f_i^{\mu \nu} \Gamma_{\mu } \Gamma_{\nu} = \frac{1}{2}\slashed{k}_i \slashed{\epsilon}_i$, where the contraction of a gamma matrix with a vector $v^\mu$ is denoted by $\slashed{v} = v^\mu \Gamma\mu$.

Now let us perform the forward limit~\cite{Caron-Huot:2010fvq} on both sides of the expansions, which involves a map for momenta of legs $-,+$: $k_{\pm}^{\mu}\to \pm \ell_{i}^{\mu}$ and the sum over their polarization states, {\it e.g.} for two external gluons:
\begin{equation} \label{eq:statesum}
\sum_{\text{states}} \epsilon_{-}^{\mu}\epsilon_{+}^{\nu}=\eta^{\mu\nu}-\frac{\ell_{i}^{\mu}q^{\nu}+\ell_{i}^{\nu}q^{\mu}}{\ell_{i}\cdot q},
\end{equation}
with $q$ to be a null reference momentum. And similarly for the external fermions we have:
\begin{equation}
    \sum_\text{states} \chi_+ \bar{\chi}_- =-\slashed{\ell_{i}},
\end{equation}
It is straightforward to show~\cite{Kosmopoulos:2020pcd} that for the expression we presented in~\eqref{eq:expansion}, we can drop the second term in~\eqref{eq:statesum} when performing the forward limit, except when contracting with $\eta_{\mu\nu}$, the $\sum_{\text{states}}\epsilon_+ \cdot \epsilon_-=D-2$, and $D$ denotes the spacetime dimension.

\begin{figure}[H]
	\centering
	\includegraphics[scale=0.45]{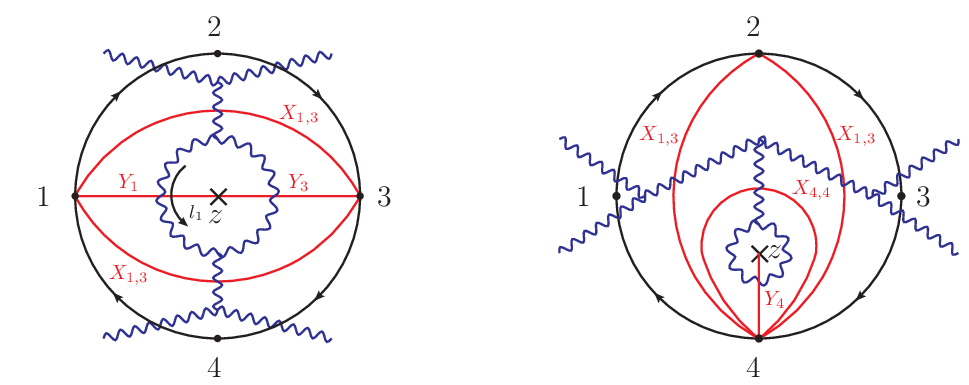}
	\caption{Triangulations of bubble and tadpole diagrams. Each curve (red line) assigned a planar variables $X_{i,j}$. We denote planar variables $X_{i,z}\equiv \ell_i^2:=(\ell+k_1+\ldots+k_{i-1})^2$ as $Y_i$.}
	\label{Fig: planarvar}
\end{figure}

For instance, on the left-hand side~\eqref{eq:expansion}, under the forward limit we simply obtain a single cut of the one-loop integrand, namely its residue on $Y_i=0$. On the right-hand side, the state sum operates solely on the prefactors, converting them into traces of the field strengths, {\it i.e.},
\begin{equation}
\begin{aligned}
   &\epsilon_{-} \cdot \epsilon_+ \xrightarrow{\fwl} D-2, \quad \\
   &\epsilon_{-}\cdot f_{\alpha_1}\cdot f_{\alpha_2}\cdots f_{\alpha_m}\cdot \epsilon_{+}\xrightarrow{\fwl}\trv(f_{\alpha_1} f_{\alpha_2}\cdots f_{\alpha_m}).
\end{aligned}
\end{equation}
where F.L. denotes the forward limit and we define $\trv(f_{\alpha_1} f_{\alpha_2}\cdots f_{\alpha_m}):=(f_{\alpha_1})_{\mu_1}^{\; \mu_2}(f_{\alpha_2})_{\mu_2}^{\; \mu_3}\cdots(f_{\alpha_m})_{\mu_m}^{\; \mu_1}$ . On the other hand, the momentum mapping affects only the mixed amplitudes, turning them into the corresponding single cuts of the scalar-loop amplitudes. Therefore the result simply reads:
\begin{equation}\label{eq:gluonloop}
\begin{aligned}
    &\mathrm{Res}_{Y_i=0} \mathcal{I}^\text{YM}_\text{gluon-loop}(1,2,\ldots,n)
    \\
    =& \sum_{m,\alpha} \trv(f_{\alpha_1} \cdots f_{\alpha_m}) \mathrm{Res}_{Y_i=0} \mathcal{I}^\text{YMS}_\text{scalar-loop}(\alpha|1,\ldots,n),
\end{aligned}
\end{equation}
where we have collected terms with identical prefactor (up to a sign) in the original sum, and the sum now spans all cyclic and reflection-inequivalent ordered subsets, {\it i.e.} $\alpha\in S_{m-1}/\mathbb{Z}_2$. Analogously, for~\eqref{eq:expansion2} the forward limit only differs in the prefactor, where we have:
\begin{equation}
\begin{aligned}
   &\bar{\chi}_{-}  \xi_+ \xrightarrow{\fwl} 2^{D/2-1},\\
   &\bar{\chi}_-  \slashed{f}_{\alpha_1} \slashed{f}_{\alpha_2} 
    \cdots \slashed{f}_{\alpha_m}  \xi_+\xrightarrow{\fwl}\trs(\slashed{f}_{\alpha_1} \slashed{f}_{\alpha_2}\cdots \slashed{f}_{\alpha_m}).
\end{aligned}
\end{equation}
where we have defined $\trs(\slashed{f}_{\alpha_1} \slashed{f}_{\alpha_2}\cdots \slashed{f}_{\alpha_m})=\left.\left(f_{\alpha_1} f_{\alpha_2} \ldots f_{\alpha_m}\right)_\gamma^\beta \delta_\beta^\gamma \right|_{\text {even }}$, and we only focus on the parity-even part of the contraction. 
The $\trs$ can also be expanded as a linear combination of $\trv$ (see equation (3.11) in~\cite{Edison:2020uzf}). We can therefore conclude that
\begin{equation}\label{eq:fermionloop}
\begin{aligned}
    &\mathrm{Res}_{Y_i=0} \mathcal{I}^\text{SYM}_\text{fermion-loop}(1,2,\ldots,n)
    \\
    =& \sum_{m,\alpha} \trs(\slashed{f}_{\alpha_1} \cdots \slashed{f}_{\alpha_m}) \mathrm{Res}_{Y_i=0} \mathcal{I}^\text{YMS}_\text{scalar-loop}(\alpha|1,\ldots,n),
\end{aligned}
\end{equation}
where $\alpha\in S_{m-1}/\mathbb{Z}_2$, and we have defined $\trv(\varnothing)=D-2$ and $\trs(\varnothing)=2^{D/2-1}$. 
\paragraph{Reconstruction of one-loop integrands}
Next we introduce the reconstruction of the complete color-ordered one-loop integrand, based on single cuts provided above via forward limits. Before proceeding, it is important to introduce the planar variables (see~\Cref{Fig: planarvar}): $X_{i,j}\equiv (k_i+ \cdots + k_{j{-}1})^2$ and $Y_i=\ell_i^2=(\ell_1+k_2+\dots+k_{i-1})^2$ (including $X_{i, i{+}1}$ and $X_{i,i}$ which can be regarded as ``regulators" for massless bubble and tadpole diagrams). Notably, there are $\frac{n(n-3)}{2}$ of $X_{i,j}$'s (excluding $X_{i, i{+}1}$ and $X_{i,i}$) and $n$ of $Y_i$'s, which is exactly the number of independent $k_i \cdot k_j$ and $\ell \cdot k_i$. The underlying geometric interpretation shows that each planar variable corresponds to a diagonal in the dual polygon with an internal puncture in the one-loop case. Furthermore, each planar trivalent diagram corresponds to a specific triangulation of this polygon. Therefore, since the planar integrand can be expressed as a summation over ordered trivalent diagrams with corresponding numerators, it can be treated as a simple rational function, where each term features a {\it monomial} denominator. This is crucial for the next step to be unambiguous.

Now we are ready to introduce the reconstruction~\footnote{Our reconstruction is similar to the recursion for scalar theories~\cite{recursion}, but we can directly apply this to gauge theories}. Let us begin with a one-loop two-point example. Note that the full integrand takes the following form:
\begin{equation}
    \mathcal{I}^\text{YM}_2=\frac{N_1}{Y_1}+\frac{N_2}{Y_2}+\frac{N_{12}}{Y_1 Y_2}+N_0.
\end{equation}
In this organization, the result is structured according to the presence of $Y_i$'s in the denominator. The numerator $N_1$ is a function independent of $Y_1$, while $N_2$ and $N_{12}$ are independent of $Y_2$ and both $Y_1,Y_2$, respectively. $N_0$ represents the boundary term with neither $Y_1$ nor $Y_2$ its denominator. This decomposition is well-defined due to the use of planar variables $Y_i$'s as our kinematic basis for $\ell\cdot k_i$'s. Then the single cut on $Y_1=0$ is given by:
\begin{equation}
    \text{Cut}_{Y_1}\equiv\mathrm{Res}_{Y_1=0} \mathcal{I}^\text{YM}_2=N_1+\frac{N_{12}}{Y_2}=N_1+\tilde{Y}_2 N_{12}, 
\end{equation}
where we have defined $\tilde{Y}_i=\frac{1}{Y_i}$. 
Similarly, we can compute the cut on $Y_2=0$, $\text{Cut}_{Y_2}=N_2+\tilde{Y}_1 N_{12}$, but there is a common term $N_{12}$ from the double residue of the integrand, {\it i.e.} $N_{12}=\mathrm{Res}_{Y_1=0,Y_{2}=0} \mathcal{I}^\text{YM}_2$. So we need to  reconstruct the full integrand with these two cuts by non-overlapping sum. A convenient way to do is that once we add a new cut, we set those $\tilde{Y}$ appeared in the original cut to zero to make the overlapping terms vanish. For example:
\begin{equation}
\begin{aligned}
\mathcal{\tilde{I}}^\text{YM}_2&=\frac{\text{Cut}_{Y_1}}{Y_{1}}+\left(\frac{\text{Cut}_{Y_2}}{Y_{2}} \Big|_{\tilde{Y}_1\to 0}\right)\\
  &= \tilde{Y}_1 N_1 +\tilde{Y}_2 N_2 +\tilde{Y}_1\tilde{Y}_2 N_{12}\,,
\end{aligned}
\end{equation}
  where we set $\tilde{Y}_1\to 0$ in the cut of $Y_{2}$ to remove the double counting of $N_{12}$. Note that $\mathcal{\tilde{I}}^\text{YM}_2$ and $\mathcal{I}^\text{YM}_2$ differ by the boundary term $N_0$, however, $N_0$ can be dropped from the integrand since it is {\it scaleless}, {\it e.g.} integrated to zero,~\footnote{Of course the full one-loop two-point integrand is scaleless, the point is the $N_0$ analog for higher multiplicity is always scaleless since it does not contain any $Y_i$'s in the denominator.} and does not contribute to the integrated result. Thus, we have reconstructed the integrand from the single cut, and this approach can be readily generalized to all multiplicities, {\it i.e.}
\begin{equation}
 \mathcal{\tilde{I}}^\text{YM}_n=\sum_{i=1}^n \frac{\text{Cut}_{Y_i}}{Y_i }\Big|_{\tilde{Y}_1,\tilde{Y}_2,\ldots,\tilde{Y}_{i-1}\to0},
\end{equation}
where we already set those $\tilde{Y}_{j}\to 0$ in the $\text{Cut}_{Y_i}$, for $j=1,\ldots,i-1$, to obtain the correct integrand without overlapping terms. All constant terms are zero/scaleless for the physical integrand, so $\mathcal{\tilde{I}}^\text{YM}_n=\mathcal{I}^\text{YM}_n$. Now applying the reconstruction procedure on both sides of~\eqref{eq:gluonloop} or~\eqref{eq:fermionloop}, crucially the prefactors on the right-hand side are never involved since they are free of $Y_i$'s, therefore we simply uplift the expansion of the single cut into the integrand level.

Note that the expansion we proposed for the integrand with quadratic propagators is valid for various gauge theories, differing only in the prefactor.  For instance, for the one-loop integrand of $\mathcal{N}=1$ SYM in $D=10$ with both fermions and gluons in the loop, we have:
\begin{equation}
    \mathcal{I}^\text{SYM}_\text{SUSY-loop}= \mathcal{I}^\text{SYM}_\text{gluon-loop}-\frac{1}{2} \mathcal{I}^\text{SYM}_\text{fermion-loop}.
\end{equation}
Therefore expansion applies with the prefactor given by:
\begin{equation}
\begin{aligned}
    &\trmax(f_{\alpha_1} f_{\alpha_2}\cdots f_{\alpha_m})\\
    =& \trv(f_{\alpha_1} f_{\alpha_2}\cdots f_{\alpha_m})-\frac{1}{2}  \trs(\slashed{f}_{\alpha_1} \slashed{f}_{\alpha_2}\cdots \slashed{f}_{\alpha_m}) .
\end{aligned}
\end{equation}

\section{Universal expansions of one-loop gluon amplitudes}
\begin{figure}[H]
	\centering
	\subfloat[{[(1,2),3,(4,5),...,n]}]{\includegraphics[align=c,height=0.15\textwidth]{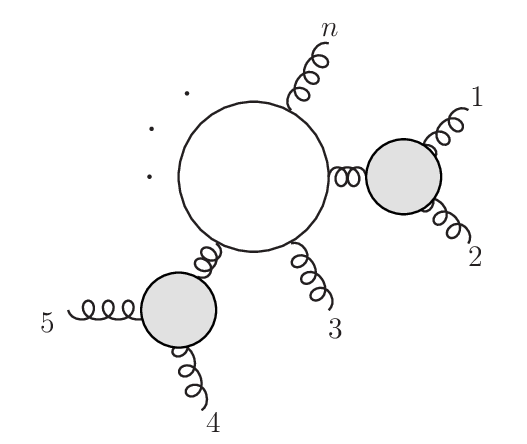}\label{fig: scalar loop examples}}\hfill
	\subfloat[bubble]{\includegraphics[align=c,height=0.15\textwidth]{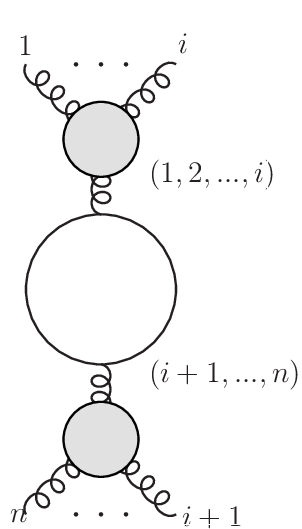}\label{fig: scalar loop bubble}}\hfill
	\subfloat[tadpole]{\includegraphics[align=c,height=0.15\textwidth]{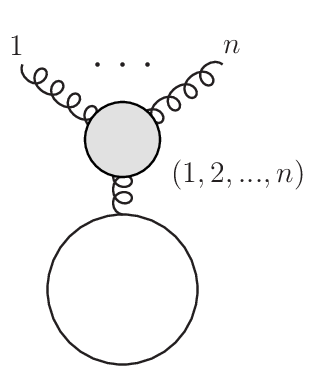}\label{fig: scalar loop tadpole}}\hfill
	\caption{Examples of  scalar loop with gluon blobs}
	\label{Fig: scalar loop examples}
\end{figure}

Let us present the universal expansions for one-loop $n$-gluon amplitudes first in pure Yang-Mills theory and we will see that a small modification leads to that in general gauge theories. The formula involves a summation over all length-$m$ cycle involving a subset of the $n$ gluons, denoted as $(\alpha_1, \alpha_2, \cdots, \alpha_m)$, for $m=0,1,\cdots, n$ (the total number is given in https://oeis.org/A116723); for pure Yang-Mills case each term is given by the Lorentz-trace of linearized field strengths, for the cycle $(\alpha_1, \cdots, \alpha_m)$ {\it e.g.} $\trv(f_{\alpha_1}f_{\alpha_2}\cdots f_{\alpha_1})$, multiplied by the amplitude of the remaining ($n{-}m$) gluons and $m$ scalars with scalar running in the loop:
\begin{equation}\label{eq: I_n^YM expansion}
{\cal I}_n^{\rm YM}=\sum_{m, \alpha\in S_{m-1}/\mathbb{Z}_2} \trv(f_{\alpha_1} \cdots f_{\alpha_m}) {\cal I}^{\rm scalar-loop}_{\alpha}\,,    
\end{equation}
where $ {\cal I}^{\rm scalar-loop}_{\alpha}\equiv \mathcal{I}^\text{YMS}_\text{scalar-loop}(\alpha|1,\ldots,n)$, and we define the special case for $m=0$ as $\trv(\emptyset)\equiv D{-}2$, and any term with $m=1$ vanishes since $\trv(f)=0$; there are ${n \choose m} \times (m{-}1)!/2$ terms for $m>2$ (and ${n \choose 2}$ terms for $m=2$), {\it e.g.}\footnote{Here we have omitted the superscript \textit{scalar-loop} in the ${\cal I}_\alpha^{\rm scalar-loop}$.} we have $2$ non-vanishing terms for $n=2$, $(D{-}2){\cal I}_{\emptyset} + \trv(f_1 f_2) {\cal I}_{1,2}$, and $5$ such terms for $n=3$: $(D{-}2){\cal I}_{\emptyset}$, $\trv(f_1 f_2){\cal I}_{1,2} + 2~\text{ perms}$ and $\trv(f_1 f_2 f_3) {\cal I}_{1,2,3}$. Let us describe these scalar-loop amplitudes in detail as they turn out to be universal building blocks for one-loop amplitudes in any gauge theory, and we first focus on the most complicated one with $m=0$, ${\cal I}_{\emptyset}$, which is the coefficient of $D{-}2$ of the one-loop amplitudes. It is given by attaching tree-level ``blobs" with external gluons to the scalar loop: the simplest diagram, the $n$-gon, has $n$ gluons attached (denoted as $[1,2,\cdots, n]$), and we have other diagrams which correspond to $(n{-}1)$-gon, $(n{-}2)$-gon,...  with subsets of gluons forming tree blobs, {\it e.g.} \Cref{fig: scalar loop examples} shows a diagram with $1,2$ in a blob, and $4,5$ in another (denoted as $[(12), 3, (45), \cdots, n]$; \Cref{fig: scalar loop tadpole,fig: scalar loop bubble} shows examples of bubble and tadpole diagrams with two blobs, {\it e.g.} $[(12\cdots i), (i{+}1 \cdots n)]$ and one blob $[(12\cdots n)]$ respectively; within each blob there are many tree-level Feynman diagrams contributing, but they do not affect the scalar-loop part. Note that ${\cal I}_\emptyset$ is gauge invariant by itself, which can be obtained as taking $D\to \infty$ limit. 

We can similarly obtain all mixed amplitudes with $m>0$ scalars where the tree blobs can also have bi-adjoint scalars (in addition to gluons) and only those diagrams consistent with the ordering $\alpha$ for these scalars contribute. 
\vspace{1em}

\begin{figure}[H]
	\centering
	\includegraphics[scale=0.5]{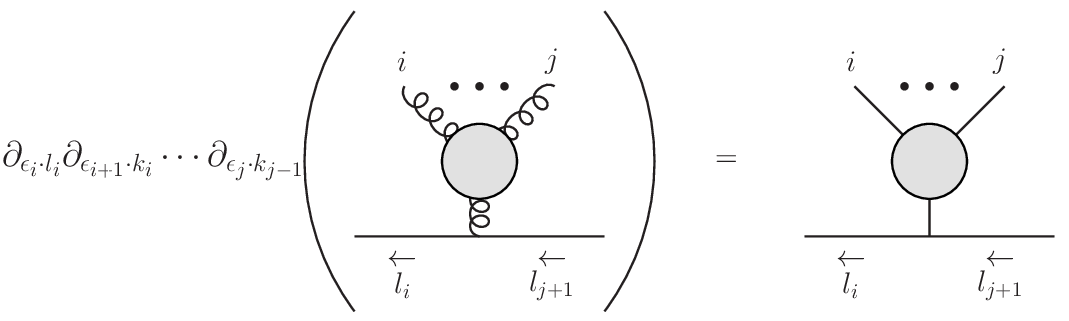}
	\caption{For the special expression of $\mathcal{I}_{\emptyset}$, each blob would succeed the transmuted operators from the tree level.}
	\label{Fig: blobop}
\end{figure}

\paragraph{One-loop integrands from transmuted operators} 
Alternatively, we can obtain the integrand for these mixed amplitudes ${\cal I}_{\alpha}$ by acting with certain differential operators on ${\cal I}_{\emptyset}$, in a way very similar to transmuted operators connecting analogous tree amplitudes~\cite{Cheung:2017ems}
\begin{equation}
\begin{aligned}
    &\mathcal{A}^{\text{YMS}}(1,\alpha,n|1,\ldots,n)=\partial_{\epsilon_{\alpha_1} \cdot k_{1}}\partial_{\epsilon_{\alpha_2} \cdot k_{\alpha_1}}\cdots\partial_{\epsilon_{\alpha_m} \cdot k_{\alpha_{m-1}}} \\
    &\left(\mathcal{A}^{\text{YMS}}(1,n|1,\ldots,n)|_{k_n\to-(k_1+\ldots k_{n-1})}\right)\,.
\end{aligned}
\label{eq: treeop}
\end{equation}
For the part with $m$ scalars $(\alpha_1, \cdots, \alpha_m)$, we need a $m$-th order differential operator linear in the derivative w.r.t. these polarizations, 

\begin{equation}
{\cal I}_\alpha^{\rm scalar-loop}=\mathcal{D}^{(m)}_\alpha {\cal I}_{\emptyset}^{\rm scalar-loop}\,,    
\end{equation}
where $\mathcal{D}^{(m)}_\alpha$ is a degree-$m$ polynomial of $\partial_{\epsilon_a \cdot k}$ for $a=\alpha_1, \cdots, \alpha_m$ and some momentum $k$.
Before going further, we point out that there is an equivalent class of $\mathcal{I}_{\emptyset}$ under the momentum conservation, and for a special expression of $\mathcal{I}_{\emptyset}$ we have the one-loop transmuted operators $\mathcal{D}^{(m)}_\alpha$ very similar to tree's one (see~\Cref{Fig: blobop}). A systematic way to get the special $\mathcal{I}_{\emptyset}$ is applying momentum conservation for every scalar loop diagram (like drawn in \Cref{Fig: scalar loop examples}) to satisfy two conditions (a)  The Lorentz product of loop momentum $l_i$ and polarization $\epsilon_j$ must be expanded to the basis $\{l_1\cdot\epsilon_1,l_2\cdot\epsilon_2,...,l_n\cdot\epsilon_n\}$;  (b) For each $\epsilon_i\cdot k_j$ the $(i,j)$ must belong to the same blob. There are ways for generating such a $\mathcal{I}_{\emptyset}$, including the reconstruction based on the forward limit.\footnote{To achieve this, we demand the tree amplitudes have solved $k_n$ like~\eqref{eq: treeop}, and trans every $l_i\cdot \epsilon_j$ to $l_i\cdot \epsilon_i$ manually} To be completely explicit, we present a $\texttt{Mathematica}$ code to generate the integrand ${\cal I}_{\emptyset}$ in an ancillary file. 

 Now we have a glimpse of that $\mathcal{D}^{(m)}_\alpha$ is a combination of some differential operator products. We define two types of derivatives
 \begin{equation}
 \partial_a^{\pm}:=\mp\partial_{\epsilon_a \cdot \ell_a}\,,\quad \partial_{(b,a)}^{\pm}:=\mp\partial_{\epsilon_b \cdot k_a}\,,
 \end{equation}
 where the signature ``$\mp$'' before the derivative is useful to describe $\mathcal{D}^{(m)}_\alpha$ because we have expanded $\trv$ to a basis so that some signatures generated from $\trv(f_{\alpha_1}f_{\alpha_2}\cdots f_{\alpha_m})=(-1)^m \trv(f_{\alpha_m}f_{\alpha_{m-1}}\cdots f_{\alpha_1})$. Then the differential operator corresponds to a tree blob with scalars $(b_1,b_2,\dots,b_q)$ (like drawn in~\Cref{Fig: blobop}) and the entire expression of ${\cal D}_\alpha^{(m)}$ are defined as
\begin{equation}
\partial_{(b_1,b_2,\dots,b_q)}^{\pm}:=\partial_{b_1}^{\pm}\partial_{(b_2,b_1)}^{\pm}\partial_{(b_3,b_2)}^{\pm}\cdots\partial_{(b_q,b_{q-1})}^{\pm},
\end{equation}
\begin{equation}\label{def: D_alpha}
    \left\{
    \begin{aligned}
        &\mathcal{D}_{ij}^{(2)}=\partial_i^{+}\partial_j^{+}\,,\\[5pt]
        &\mathcal{D}_{\alpha}^{(m>2)}=\sum_{\mathcal{G}\subset\Gamma[\alpha]}\prod_{B\in \mathcal{G}}\partial_B^{+}
	+\sum_{\mathcal{G}\subset\Gamma[\tilde{\alpha}]}\prod_{B\in \mathcal{G}}\partial_{B}^{-},
    \end{aligned}
    \right.
\end{equation}
where $\tilde{\alpha}$ denotes the reversed $\alpha$ and $\Gamma[\alpha]$ is a set of scalar loop diagrams with scalar blobs, referred to as scalar skeleton diagrams. Each scalar skeleton diagram in $\Gamma[\alpha]$ can be denoted as a partition of the cycle $\alpha$, like $[\alpha_1,(\alpha_2,\alpha_3),\ldots]\,, [(\alpha_i,\alpha_{i+1}),\alpha_{i+2},\ldots]\,,\ldots$. And we demand the scalar skeleton diagrams in $\Gamma[\alpha]$ could be recovered to the canonical ordering through some permutations on every blob, or it's inconsistent with $\mathcal{I}^\text{YMS}_\text{scalar-loop}(\alpha|12\dots n)$.

\begin{figure}[H]
	\centering
	\subfloat[{$[1,(3,2)]\in\Gamma[132]$}]{\includegraphics[align=c,scale=0.45]{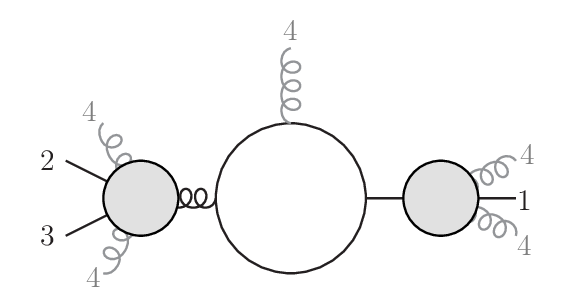}\label{fig: scalarskeletonbubble}}\hfill
	\subfloat[{$[(1,3,2)]\in\Gamma[132]$}]{\includegraphics[align=c,scale=0.45]{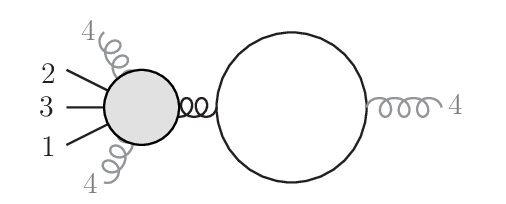}\label{fig: scalarskeletontadpole}}\hfill
    \\
    \subfloat[{$[1,3,2]\notin\Gamma[132]$}]{\includegraphics[align=c,scale=0.45]{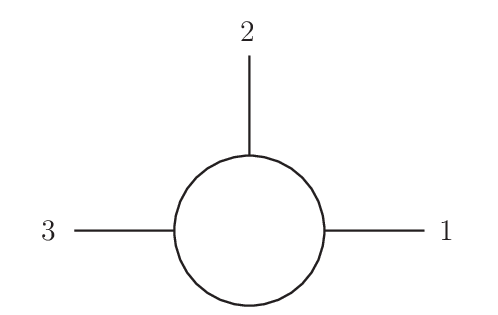}}\hfill
	\subfloat[{$[2,3,1]\in\Gamma[231]$}]{\includegraphics[align=c,scale=0.45]{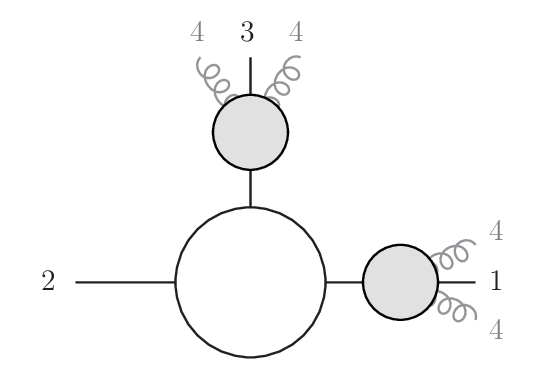}}\hfill
	\caption{Examples of scalar skeleton diagrams (draw in black lines). Each scalar skeleton diagram is a mixed loop diagram that deletes external gluons in all blobs. The gray lines denote the possible positions of deleted gluons.}
	\label{Fig: skeleton}
\end{figure}

There are three types of scalar skeleton diagrams for $n=4$ and $\alpha=132$ (see~\Cref{Fig: skeleton}), and we have $\Gamma[132]=\{[(1,3,2)],[1,(3,2)],[(1,3),2],[(2,1),3]\}$, $\Gamma[231]=\{[(2,3,1)],[2,(3,1)],[(2,3),1],[(1,2),3],[2,3,1]\}$, plug $\Gamma[132]$ and $\Gamma[231]$ into~\eqref{def: D_alpha} and we have
\begin{equation}\label{eq: D132}
	\begin{aligned}
		\mathcal{D}_{132}^{(3)}=&(\partial_{1}^{+}\partial_{(3,2)}^{+}+\partial_{(2,3)}^{-}\partial_{1}^{-}+{\rm cyclic})\\
		&+\partial_{(1,3,2)}^{+}+\partial_{(2,3,1)}^{-}+\partial_{2}^{-}\partial_{3}^{-}\partial_{1}^{-}.
	\end{aligned}
\end{equation}

 For each ``scalarblob operator'' $\partial_B^{\pm}$, there is an equivalent class, which helps us to relate $\partial_B^{+}$ and $\partial_B^{-}$,
\begin{equation}\label{eq: equivalence of operator}
	\partial_{(b_1,b_2,\dots,b_m)}^{\pm}\equiv \partial_{(b_1,b_2)}^{\mp}\cdots\partial_{(b_{i-1},b_i)}^{\mp}\partial_{(b_i,b_{i+1},\dots,b_m)}^{\pm}\,,
\end{equation}
where $i$ is arbitrary and we define $\partial_{(b_m,b_m)}^{\pm}=\partial_{b_m}^{\pm}$. The~\eqref{eq: equivalence of operator} trans $\partial^{\pm}$ to products of $\partial^{+}$ and $\partial^{-}$, for example, $\partial_{(2,3,1)}^{-}\equiv \partial_{(2,3)}^{+}\partial_{(3,1)}^{+}\partial_{(1)}^{-}=-\partial_{(1,3,2)}^{+}$. Combine~\eqref{def: D_alpha} with~\eqref{eq: equivalence of operator} and one can directly derive that the two tadpole scalar skeleton diagrams (see~\Cref{fig: scalarskeletontadpole}) from $\Gamma[b_1 b_2 ... b_q]$ and $\Gamma[b_q b_{q-1} ... b_1]$ would cancel each other,
\begin{equation}
    \partial_{(b_1,b_2,...,b_q)}^{+}\equiv \partial_{(b_1,b_2)}^{-}\dots\partial_{(b_{q-1},b_{q})}^{-}\partial_{b_q}^{+}=-\partial_{(b_q,b_{q-1},...,b_1)}^{-}\,,
\end{equation} 
and the two bubble scalar skeleton diagrams (see~\Cref{fig: scalarskeletonbubble}) from $\Gamma[b_1 b_2 ... b_q]$ and $\Gamma[b_q b_{q-1} ... b_1]$ would equal each other
\begin{equation}
    \partial_{(b_1,...,b_p)}^{+}\partial_{(b_{p+1},...,b_q)}^{+}=\partial_{(b_q,...,b_{p+1})}^{-}\partial_{(b_{p},...,b_1)}^{-}\,.
\end{equation}
Now we can simplify~\eqref{eq: D132} by transforming $\partial^{-}$ to $\partial^{+}$  
\begin{equation}\label{eq: simple D_132}
 \mathcal{D}_{132}^{(3)}=(2\partial_{1}^{+}\partial_{(3,2)}^{+}+{\rm cyclic})-\partial_{1}^{+}\partial_{3}^{+}\partial_{2}^{+}\,.
\end{equation}
In the ancillary $\texttt{Mathematica}$ file, there is a command $\texttt{Bloboperator}$ that can generate any $\mathcal{D}_\alpha^{(m)}$ in~\eqref{def: D_alpha} with transforming $\partial^{-}$ to $\partial^{+}$ like~\eqref{eq: simple D_132}. With the help of those differential operators, the $\mathcal{I}_n^{\text{YM}}$ can be rebuilt by~\eqref{eq: I_n^YM expansion}. We implement those in the $\texttt{Mathematica}$ file, to help us to generate the explicit $\mathcal{I}_n^{\text{YM}}$ results.
\vspace{1em}

\paragraph{General gauge theories and SYM}
We have essentially the same formula for the parity-even part of one-loop amplitudes in general gauge theories where in addition to gluons, fermions and/or scalars can also run in the loop. Based on previous results of tree amplitudes with $n$ gluons and an additional pair of particles in the forward limit~\cite{Edison:2020uzf}, all we need to do is to replace the Lorentz trace by more general combinations: with ${\bf n}_v, {\bf n}_f, {\bf n}_s$ species of vectors, Weyl fermions and scalars in $D$ dimensions. 

The one-loop integrand for general gauge theory is, 
\begin{equation}
{\cal I}_n=\sum_{m, \alpha\in S_{m-1}/\mathbb{Z}_2} {\cal T}
_{\alpha_1, \cdots \alpha_m}{\cal I}^{\rm scalar-loop}_{\alpha}\,, \end{equation}
where we replace the Lorentz trace by
\begin{equation}
{\cal T}
_{\alpha_1, \cdots \alpha_m}={\bf n}_v {\rm tr}_V(\alpha_1 \cdots \alpha_m)-\frac{{\bf n}_f}{2} {\rm tr}_S(\alpha_1 \cdots \alpha_m)\,,
\end{equation}
for $m>0$ where we renamed ${\rm tr}_V(12\cdots)\equiv {\rm tr}_{V}(f_1 f_2 \cdots)$ the Lorentz (vector) trace and also the spinor trace ${\rm tr}_S$ defined in the section~\ref{sec:review}, and see also~\cite{Edison:2020uzf,Edison:2022jln} ; in the special case of $m=0$, ${\cal T}_{\emptyset}=(D{-}2) {\bf n}_v + {\bf n}_s-2^{(D{-}2)/2} \frac{{\bf n}_f}2$ which counts the number of on-shell (bosonic and fermionic) degrees of freedom (and it is the only place ${\bf n}_s$ appears). 

In particular, if we are interested in maximally-supersymmetric theories, such as super-Yang-Mills (SYM) in $D=10$, we have $({\bf n}_f,{\bf n}_s)=(1,0){\bf n}_v$ and similarly we have $({\bf n}_f, {\bf n}_s)=(4,0) {\bf n}_v$ in $D=6$ and $({\bf n}_f, {\bf n}_s)=(8,6) {\bf n}_v$ in $D=4$. As discussed in~\cite{Edison:2020uzf,Edison:2022jln}, for $m<4$ this combination is proportional to ${\bf n}_v-2^{D/2-5} {\bf n}_f$ which vanishes for all these cases. For example in $D=10$, ${\cal T}_{\alpha}=0$ $m<4$ and for $m=4$ it is proportional to the famous $t_8$ tensor (defined in below):
\begin{equation}
{\rm tr}_V-\frac 1 2 {\rm tr}_S=\frac 1 2 [{\rm tr}_V(1,2,3,4)-\frac{1}{4} {\rm tr}_V(1,2){\rm tr}_V(3,4)+ {\rm cyc.}]
\end{equation}
This is a remarkable cancellation which leads to the correct power-counting for SYM amplitudes: all triangles, bubbles {\it etc.} in ${\cal I}_{\alpha}$ cancel in the full integrand and we have $\ell^{r{-}4}$ in the numerator of an $r$-gon ($r\geq 4$), {\it e.g.} the box numerator is independent of $\ell$.

\section{Conclusions and outlook}
We have proposed new universal expansions for one-loop amplitudes in general gauge theories, expressing these amplitudes in terms of scalar-loop amplitudes and gauge-invariant prefactors. By applying the surface-inspired reconstruction~\cite{recursion} to the forward limits of tree amplitudes, we derived integrands with manifestly universal structures that are applicable to both pure Yang-Mills and super Yang-Mills theories. Our result is tailored for the integrands with standard quadratic propagator, distinguishing it from approaches based on linear propagators. We also explore the general structure of our expansion in the large $D$ limit, and express the remaining simpler scalar amplitudes as the result of differential operators acting on the large $D$ component.

Our preliminary studies have opened up several new avenues for investigations. The nature of surface-inspired method suggests that our results can be generalized to higher-loop orders, a process that is relatively straightforward for pure Yang-Mills theory. However, for super Yang-Mills theories, even at the two-loop level, a more comprehensive understanding of one-loop amplitudes with two external fermions is required. Addressing this challenge remains an important task for future research. It would also be interesting to investigate the structure of the leading large $D$ limit at higher loops, {\it i.e.} the $D^L$ component at $L$-loop. This term, corresponding to scalar loop integrands with $n$ external gluons, could serve as a crucial building block for constructing the complete gluon loop integrands. Such an investigation may provide deeper insights in simplifying loop-level computations. Finally, it would be desired to obtain the BCJ numerators even at the one-loop level and establishing a connection to recent developments based on worldsheet methods~\cite{Edison:2021ebi,Feng:2022wee,Rodriguez:2023qir,Dong:2023stt,Zhang:2024yfp,Balli:2024wje,Geyer:2024oeu,Xie:2024pro}.

\section*{acknowledgments}
\begin{acknowledgments}
We thank Nima Arkani-Hamed, Carolina Figueiredo, Henrik Johansson, Oliver Schlotterer, Fei Teng and Yong Zhang for stimulating discussions and collaborations on related projects. J.D. thanks Uppsala University for hospitality during his visit. The work of Q.C. is supported by the National Natural Science Foundation of China under Grant No. 123B2075. The work of S.H. has been supported by the National Natural Science Foundation of China under Grant No. 12225510, 11935013, 12047503, 12247103, and by the New Cornerstone Science Foundation through the XPLORER PRIZE. 
\end{acknowledgments}

\bibliographystyle{apsrev4-1}
\bibliography{Refs}


\end{document}